\documentclass[amssymb,amsmath,aps,showpacs,floatfix,nofootinbib,showpacs,12pt]{revtex4}
\usepackage{epsfig}
\usepackage{color}
\usepackage{graphicx}

\begin{document}
\title{Parameters in a Class of Leptophilic Dark Matter Models from PAMELA, ATIC and FERMI
}

\author{$^{1,2}$Xiao-Jun Bi, $^{1,3}$Xiao-Gang He, $^2$Qiang Yuan}
\affiliation{
$^1$Center for High Energy Physics, Peking University, Beijing 100871\\
$^2$Laboratory of Particle Astrophysics, Institute of High Energy Physics,Chinese Academy of Sciences, Beijing 100049\\
$^3$Department of Physics, Center for Theoretical
Sciences, and LeCospa Center, National Taiwan University,
Taipei}

\begin{abstract}

In this work we study a class of leptophilic dark matter models,
where the dark matter interacts with the standard model particles
via the $U(1)_{L_i-L_j}$ gauge boson, to explain the $e^{\pm}$
excess in cosmic rays observed by ATIC and PAMELA experiments, and
more recently by Fermi experiment. There are three types of
$U(1)_{L_i-L_j}$ models: a) $U(1)_{L_e - L_\mu}$, b) $U(1)_{L_e -
L_{\tau}}$, and c) $U(1)_ {L_e-L_\tau}$. Although ATIC or Fermi
data is consistent with PAMELA data separately, ATIC and Fermi
data do not agree with each other. We therefore aim to identify which of the three
models can explain which data set better. We find that models a)
and b) can give correct dark matter relic density and explain the
ATIC and PAMELA data simultaneously recur to the Breit-Wigner
enhancement.  Whereas model c) with a
larger $Z^\prime$ mass can explain Fermi and PAMELA data simultaneously. In all cases
the model parameters are restricted to narrow regions. Future improved
data will decide which set of data are correct and also help to
decide the correct dark matter model.

\end{abstract}


\maketitle

\section{Introduction}

Recently, the ATIC and PPB-BETS balloon experiments have observed
excess in the $e^+ + e^-$ energy spectrum between 300 and 800 GeV
\cite{Chang:2008zz, Torii:2008xu}. The PAMELA collaboration has
also reported excesses in the positron fraction from 10 to $\sim
100$ GeV, but shown no excess for the antiproton data
\cite{Adriani:2008zr, Adriani:2008zq} compared with the prediction
in the cosmic ray physics.  These results are compatible with the
previous HEAT and AMS01 experiments (e.g.,
\cite{Barwick:1997ig,Aguilar:2007yf,ATIC-2}) with higher
precision.  Newly published result from Fermi also shows an excess
at the $e^+ + e^-$ energy spectrum above the background of the
conventional cosmic ray model. However, it shows a softer spectrum
than ATIC \cite{fermi}. The excesses may be explained by
astrophysical processes, for instance the nearby pulsars
\cite{Hooper:2008kg,Yuksel:2008rf,Profumo:2008ms} or photon-cosmic
ray interactions \cite{Hu:2009bc}, or due to annihilation or decay
of dark matter (DM) particles in our Galactic neighbourhood
predominately into leptons (e.g.,
\cite{darkmatter,Cirelli:2008pk,ArkaniHamed:2008qn,
Pospelov:2008jd,Yin:2008bs,Fox:2008kb,baek}).  Similar explanation
can also account for the excess observed at Fermi by assuming a
softer injection electron spectrum from these sources \cite{berg}.
PAMELA indicate that the dark matter is hadrophobic or
leptophilic\cite{Fox:2008kb}. In this work we show that  gauged
$U(1)_{L_i - L_j}$ models proposed some time ago\cite{model} in
searching for simple $Z^\prime$ can naturally explain the excess
in electron/positron spectrum through the DM annihilation
mechanism. Here $L_i$ and $L_j$ are one of the three family lepton
numbers with $i\neq j$.

 In constructing dark matter models, one should note that
although ATIC or Fermi data is consistent with PAMELA data
separately, ATIC and Fermi data do not agree with each other.
Improved data are needed to decide which ones are correct and to
distinguish different dark matter models. In the models we are
considering, there are three different ways to gauge the lepton
number differences: a) $U(1)_{L_e - L_\mu}$, b) $U(1)_{L_e -
L_{\tau}}$, and c) $U(1)_ {L_e-L_\tau}$. Each of them has
different features. Our aim in this work is to identify which of
the three models can explain which data set better.

The $Z^\prime$ in the gauged $U(1)_{L_i-L_j}$ is leptophilic which
only interacts with standard model (SM) leptons. If the $Z^\prime$
also interacts with DM\cite{Cirelli:2008pk,baek}, it will be the
main mediator for DM annihilation with final product dominated by
leptons offering possible explanation to  electron/positron excess
without anti-proton excess at PAMELA data.

Another requirement for DM annihilation to account for the
positron excess is that the annihilation rate $\langle \sigma
v\rangle$ determining the electron/positron spectrum should be
much larger than that determined from the usual thermal DM relic
density. The enhancement factor, usually referred as the boost
factor, is in the range of $100\sim 1000$. Several mechanisms have
been proposed to produce a large boost factor, including the DM
substructures, DM nonthermal production mechanism
\cite{nonthermal}, the Sommerfeld effect
\cite{Cirelli:2008pk,Sommerfeld,ArkaniHamed:2008qn,Pospelov:2008jd}
and the Breit-Wigner resonance enhancement effect \cite{breit}.
The detailed calculation based on the N-body simulation shows that
the boost factor from DM substructures is generally less than
$\sim 10$ \cite{Lavalle:1900wn,Lavalle:2008zb}. In order for the
Sommerfeld effect to be effective, the mediating particle needs to
be very light to allow long range interaction between DM
particles. This mechanism in our case will result in a light
$Z^\prime$ boson. However, there are very tight constraints on the
coupling constant for such light $Z^\prime$. Instead the
Breit-Wigner resonance enhancement mechanism works very well in
our model if the $Z^\prime$ mass is about two times of the dark
matter mass. From the relic density and ATIC, Fermi and PAMELA
data we can also constrain the $U(1)_{L_i-L_j}$ charge of DM.
There is no need to have very different $U(1)_{L_i-L_j}$ charges
of DM and SM leptons as in \cite{Fox:2008kb}. Further, the
determined $U(1)_{L_i-L_j}$ gauge coupling strength is consistent
with all constraints from LEP, $(g-2)_\mu$ and other data.

The ATIC data also show a sharp falling at about $600$ GeV in the
electron/positron energy spectrum. We find such a feature needs
substantial electron component as the DM annihilation products
with the electron and positron pairs have a fixed energy from DM
annihilation. Additional electron/positron energy spectrum with
lower energy from secondary decays will then help to enhance
positron  with lower energies. We find that the gauged $U(1)_{L_e
- L_\mu}$ (model a)) and $U(1)_{L_e - L_\tau}$ (model b)) can give
excellent fit to the ATIC data while the $U(1)_{L_\mu - L_\tau}$
can not. On the contrary, Fermi shows a much softer electron
spectrum, which does not favor the initial electron component. We
find that the gauged $U(1)_{L_\mu - L_\tau}$ (model c)) gives an
excellent fit to the Fermi, PAMELA and HESS data.

The paper is organized as following: in Sec. II we give a
brief introduction of the model. In Sec. III we introduce the
Breit-Wigner mechanism and the numerical results of relic density
and boost factor in our model. Then we show the electron/positron
spectrum of our model in Sec. IV. Finally we give discussions and
conclusions in Sec. V.

\section{The Model}

One of the following global symmetries in the SM can be gauged without gauge anomalies\cite{model}
\[
L_e - L_\mu , \ \  L_e - L_\tau , \ \ L_\mu - L_\tau\ .
\]
The gauge boson $Z^\prime$ resulting from one of the above models
has the desired leptophilic couplings.
At the tree-level the
$Z^\prime$ only couples to one of the pairs $e$ and $\mu$, $e$ and
$\tau$, and $\mu$ and $\tau$. We will use $Y^\prime$ to indicate
the quantum numbers for one of the above three possibilities,
$Y^\prime = L_i - L_j$. If the $Z^\prime$ in one of these models
is the messenger mediating dark matter annihilation, the resulting
final states are mainly leptonic states which can lead to
electron/positron excess observed in cosmic rays. It is then
desirable to have the $Z^\prime$ to couple to dark matter\cite{Cirelli:2008pk,baek}. We
therefore introduce a new vector-like fermion
$\psi$ with a non-trivial $Y^\prime$ number $a$. The reason for
the dark matter being vector-like is to make sure that the theory
does not have gauge anomaly required for self consistency. The
$Z^\prime$ boson can develop a finite mass from spontaneous
$U(1)_{L_i-L_j}$ symmetry breaking of a scalar $S$ with a
non-trivial charge $Y^\prime = b$.  With the new particles $Z^\prime$, $S$ and $\psi$
in the model, addition term $L_{new}$ have to be added to the
Lagrangian beside the SM one $L_{SM}$ with
\begin{eqnarray}
L_{new} &=& - {1\over 4} Z^{'\mu\nu} Z^\prime_{\mu\nu} + \sum_l \bar l \gamma^\mu (- g^\prime Y_l^\prime  Z^\prime_\mu)l
+ \bar \psi [\gamma^\mu (i\partial_\mu - a g^\prime Z^\prime_\mu) - m_\psi]\psi \nonumber\\
&+ & (D_\mu S)^\dagger (D^\mu S) + \mu^2_S S^\dagger S + \lambda_S (S^\dagger S)^2 + \lambda_{SH} (S^\dagger S) H^\dagger H\;,
\end{eqnarray}
where $l$ is summed over the SM leptons. $H$ is the usual SM Higgs
doublet.

The $Z^\prime$ coupling to fermions are given by
\begin{eqnarray}
L = - g'(a \bar \psi\gamma^\mu \psi + \bar l_i \gamma^\mu l_i - \bar l_j \gamma^\mu l_j
+ \bar \nu_i \gamma^\mu L \nu_i - \bar \nu_j \gamma^\mu L \nu_j) Z^\prime_\mu\;.
\end{eqnarray}
Note that the $Z^\prime$ coupling to leptons are flavor diagonal. There is no tree level flavor changing neutral current induced by $Z^\prime$.

After $S$ and $H$ develop non-vanishing vacuum expectation values
(vev) $v_S$ and $v$, the physical components from $S$ and $H$ can
be written as $(v_S+s)/\sqrt{2}$ and $(v+h)/\sqrt{2}$,
respectively. The non-zero $v_S$ will induce a non-zero $Z^\prime$
mass given by: $m^2_{Z^\prime} = b^2 g^{\prime 2} v_S^2$. A non-zero $v_S$ together
with the non-zero $v$ will induce mixing between $s$ and $h$
with the mixing parameter proportional to $\lambda_{SH} v v_S$.
This mixing will change the masses $m_s$ and $m_h$ of $s$ and $h$
in the limit of without mixing. Since we will require $m_{Z^\prime}$
to be much larger than the $Z$ mass, this implies that $m_s$ is
also much larger than $m_h$. The mixing will reduce the usual
Higgs mass $m_h$. However, since the parameter $\lambda_{SH}$ is
not fixed, if it is small enough the reduction in Higgs mass can
be neglected. In any case, the effects of the mixing and also other term in the Higgs potential involving $S$ and $H$ will
not affect our discussions in the following. We will not discuss them further here.

The relic density of the dark matter is controlled
by annihilation of $\bar \psi \psi \to Z^{\prime *} \to l_i \bar l_i + \nu_i \bar \nu_i$.
The annihilation rate of dark matter $\sigma v$, with lepton masses neglected and summed
over the two types of charged leptons and neutrinos, is given by
\begin{eqnarray}
\sigma v = {3\over \pi} {a^2 g^{\prime 4}m^2_\psi\over (s -
m^2_{Z^\prime})^2 + \Gamma^2_{Z^\prime} m^2_{Z^\prime}}\; ,\label{cs}
\end{eqnarray}
where $v$ is the relative velocity of the two annihilating dark matter and  $s$ is the total dark matter pair energy squared in the
center of mass frame. $\Gamma_{Z^\prime}$ is the decay width of
the $Z^\prime$ boson. If the $Z^\prime$ mass is below the $\bar
\psi \psi$ threshold which we will assume, the dominant decay
modes of $Z^\prime$ are $Z^\prime \to \bar l_i l_i + \bar \nu_i
\nu_i$, and $\Gamma_{Z^\prime}$ is given by, neglecting lepton
masses
\begin{eqnarray}
\Gamma_{Z^\prime} = {3 g^{\prime 2}\over 12 \pi} m_{Z^\prime}\;.\label{dr}
\end{eqnarray}

In Eqs. (\ref{cs}) and (\ref{dr}), we have assumed that there are
only left-handed light neutrinos. If there are light right-handed
neutrinos to pair up with left-handed neutrinos to form Dirac
neutrinos, the factor 3 in these equations should be changed to 4.

\section{The Breit-Wigner enhancement and boost factor}

Since the relic density of DM is determined by the annihilation
rate, the model parameters are thus constrained. The same
parameters will also determine the annihilation rate producing the
electron/positron excess observed today, which requires a much
larger annihilation rate. A boost factor in the range 100 to 1000
is necessary.
We find that Breit-Wigner resonance enhancement mechanism
works very well in our models if the $Z^\prime$ boson mass is about
two times of the dark matter mass.

The boost factor in this case
comes from the fact that since the $Z'$ mass $m_{Z'}$ is close to two times of the dark matter mass  $m_\psi$, the
annihilate rate is close to the resonant point and is very sensitive to the thermal kinetic energy of dark matter.
To see this let us rewrite the annihilation rate into a pair of charged leptons as
\begin{eqnarray}
\sigma v  = {a^2 g^{\prime 4}\over 16 \pi m^2_\psi} {1\over (\delta + v^2/4)^2 + \gamma^2}\;,
\end{eqnarray}
 where we have used the non-relativistic limit of
$s = 4 m^2_\psi + m^2_\psi v^2$, with $\delta$ and $\gamma$
defined as $m^2_{Z^\prime} = 4 m^2_\psi (1-\delta)$, and $\gamma^2
= \Gamma^2_{Z^\prime}(1-\delta)/4 m^2_\psi$.

For thermal dark matter, the velocity $v^2$ is proportional to the
thermal energy of dark matter. It is clear that for small enough
$\delta$ and $\gamma$, the annihilation rate is very sensitive to
the thermal energy and therefore the thermal temperature T. At
lower dark matter thermal energies, the annihilation rate is
enhanced compared with that at higher temperature. This results in
a very different picture of dark matter annihilation than the case
for the usual non-resonant annihilation where the annihilation
rate is not sensitive to dark matter thermal energies. The
annihilation process does not freeze out even after the usual
``freeze out'' time in the non-resonant annihilation case due to the
enhanced annihilation rate at lower energies. To produce the
observed dark matter relic density, the annihilation rate at zero
temperature is required to be larger than the usual one, and
therefor a boost factor. With appropriate $\delta$ and $\gamma$, a
large enough boost factor can be produced.

For a detailed discussion, a precise form for
the thermally averaged annihilation rate should be used which can be
written as\cite{breit}
\begin{equation}
\langle \sigma v\rangle=\frac{1}{n_{EQ}^2}\frac{m_{\psi}}{64\pi^4 x}
\int_{4m_{\psi}^2}^{\infty} \hat{\sigma}(s)\sqrt{s}K_1\left(
\frac{x\sqrt{s}}{m_{\psi}}\right){\rm d}s,
\end{equation}
with
\begin{eqnarray}
n_{EQ} &=& \frac{g_i}{2\pi^2}\frac{m_{\psi}^3}{x}K_2(x),\\
\hat{\sigma}(s) &=& 2g_i^2m_{\psi}\sqrt{s-4m_{\psi}^2}\cdot \sigma v,
\end{eqnarray}
where $g_i$ is the internal degrees of freedom of DM particle
which is equal to 4 for a vector fermion, $K_1(x)$ and $K_2(x)$
are the modified Bessel functions of the second type.

\begin{figure}[!ht]
\includegraphics[scale=0.5]{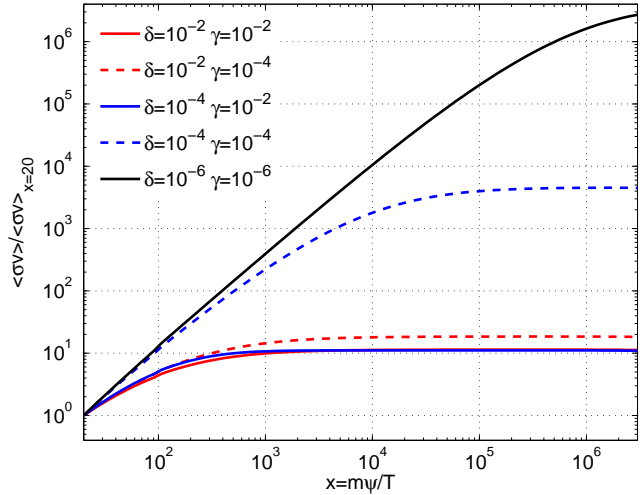}
\caption{The Breit-Wigner enhanced relative cross section
$\langle\sigma v\rangle/\langle\sigma v\rangle_{x=20}$ as a
function of time $x$.} \label{boost}
\end{figure}

We plot in Fig. \ref{boost} the thermally averaged annihilation
rate $\langle \sigma v\rangle$ as a function of cosmic
time $x\equiv m_{\psi}/T$ for several values of parameters
$\gamma$ and $\delta$. The annihilation rate when ``freeze out''
at $x\approx 20$ is adopted as a normalization to illustrate the
enhancement effect today.

\begin{figure}[!ht]
\includegraphics[scale=0.5]{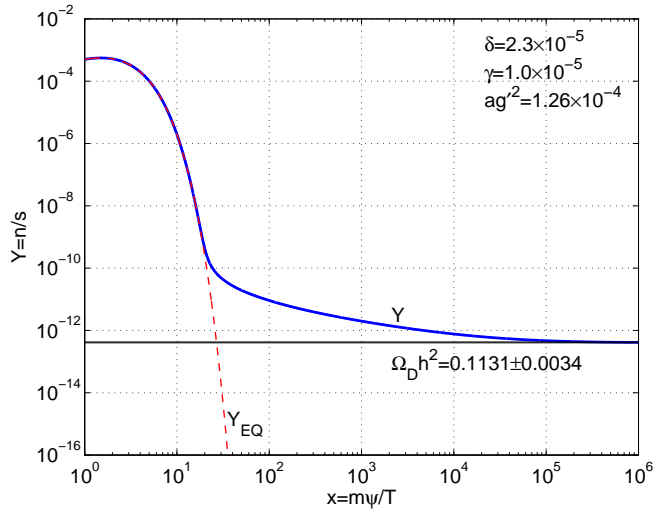}
\caption{The evolution of DM abundance $Y$ as a function of $x$,
compared with the equilibrium abundance $Y_{EQ}\propto
x^{3/2}e^{-x}$ and the experimental measurements $\Omega_Dh^2 =
0.1131\pm0.0034$ \cite{Komatsu:2008hk}.} \label{abun}
\end{figure}

To precisely determine the parameters we solve the standard Boltzmann
equation of the decoupling process numerically.
Fig. \ref{abun} shows an example of the
evolution of DM abundance $Y=n/s$ with $n$ and $s$ the number
density and entropy density respectively. The parameter
$ag^{\prime2}$ is adjusted to make sure that today's DM abundance is correct
$Y(x=3\times 10^6)=Y_0 \equiv {\Omega_Dh^2}/{2.8\times
10^8(m_{\psi}/{\rm GeV})}$\; .

\section{ The electron/positron spectrum}

 We now discuss the DM annihilation produced electron/positron spectrum in the
$U_{L_i-L_j}$ DM models for all the three possibilities discussed
before: a) $L_e - L_\mu$; b) $L_e - L_\tau$; and c) $L_\mu -
L_\tau$. In all these cases, 2/3 of the DM annihilate into charged
lepton pairs and 1/3 into neutrino pairs, therefore the required
boost factor should be 1.5 times of the case for DM only
annihilate into charged lepton pairs.

After the electron/positron pairs produced by DM annihilation they
propagate diffusively in the Galaxy due to the scattering with
random magnetic field~\cite{Gaisser:1990vg}. The propagation
processes in the Galaxy is calculated numerically in order to
compare with data measured at the Earth. The interactions with
interstellar medium, mainly the synchrotron radiation and inverse
Compton scattering processes, will lead to energy losses of the
primary electrons and positrons. In addition, the overall
convection driven by the Galactic wind and reacceleration due to
the interstellar shock will also affect the electron spectrum. In
this work we solve the propagation equation numerically adopting
the GALPROP package \cite{galprop}.

\begin{figure}[!ht]
\includegraphics[width=0.45\columnwidth]{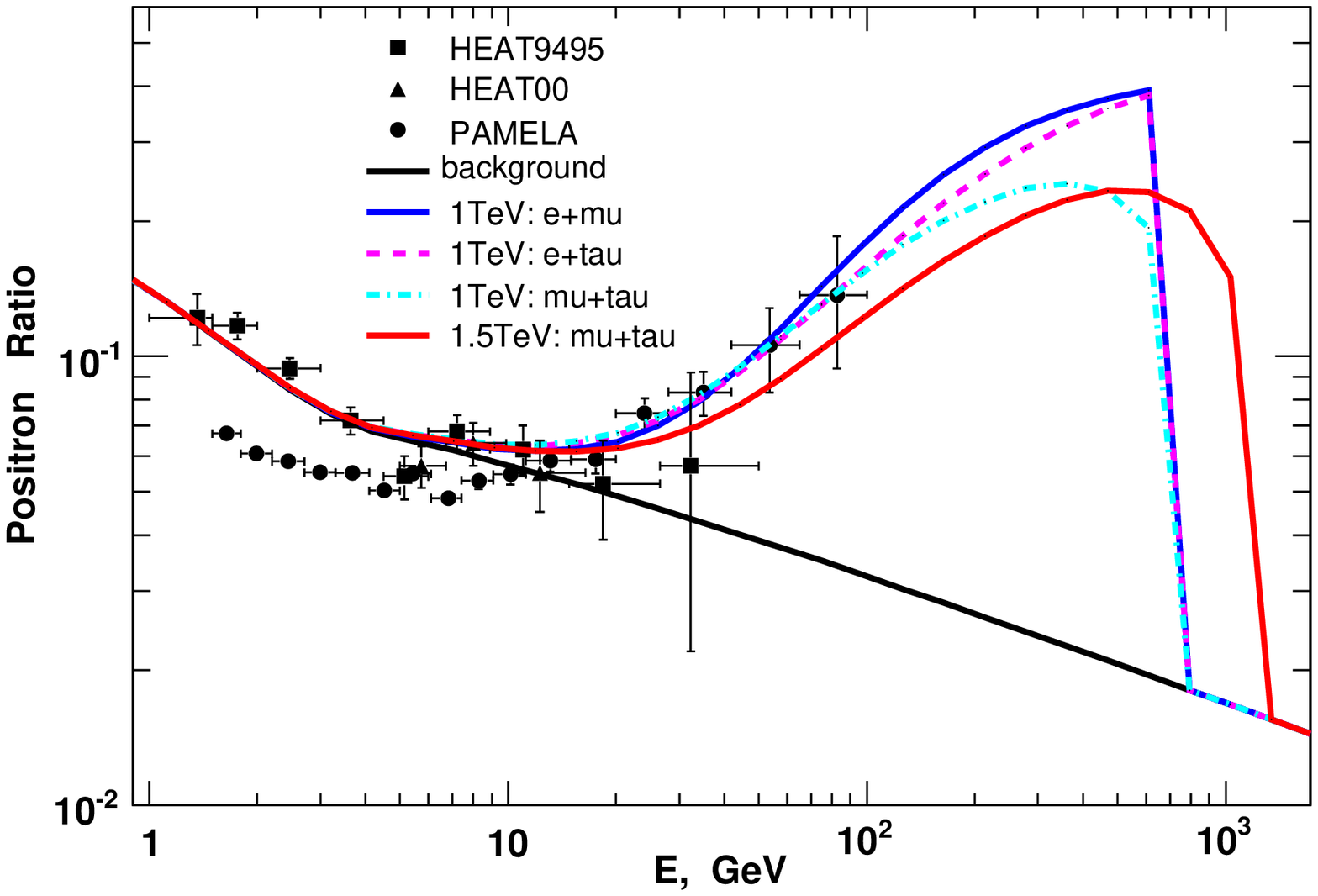}
\includegraphics[width=0.45\columnwidth]{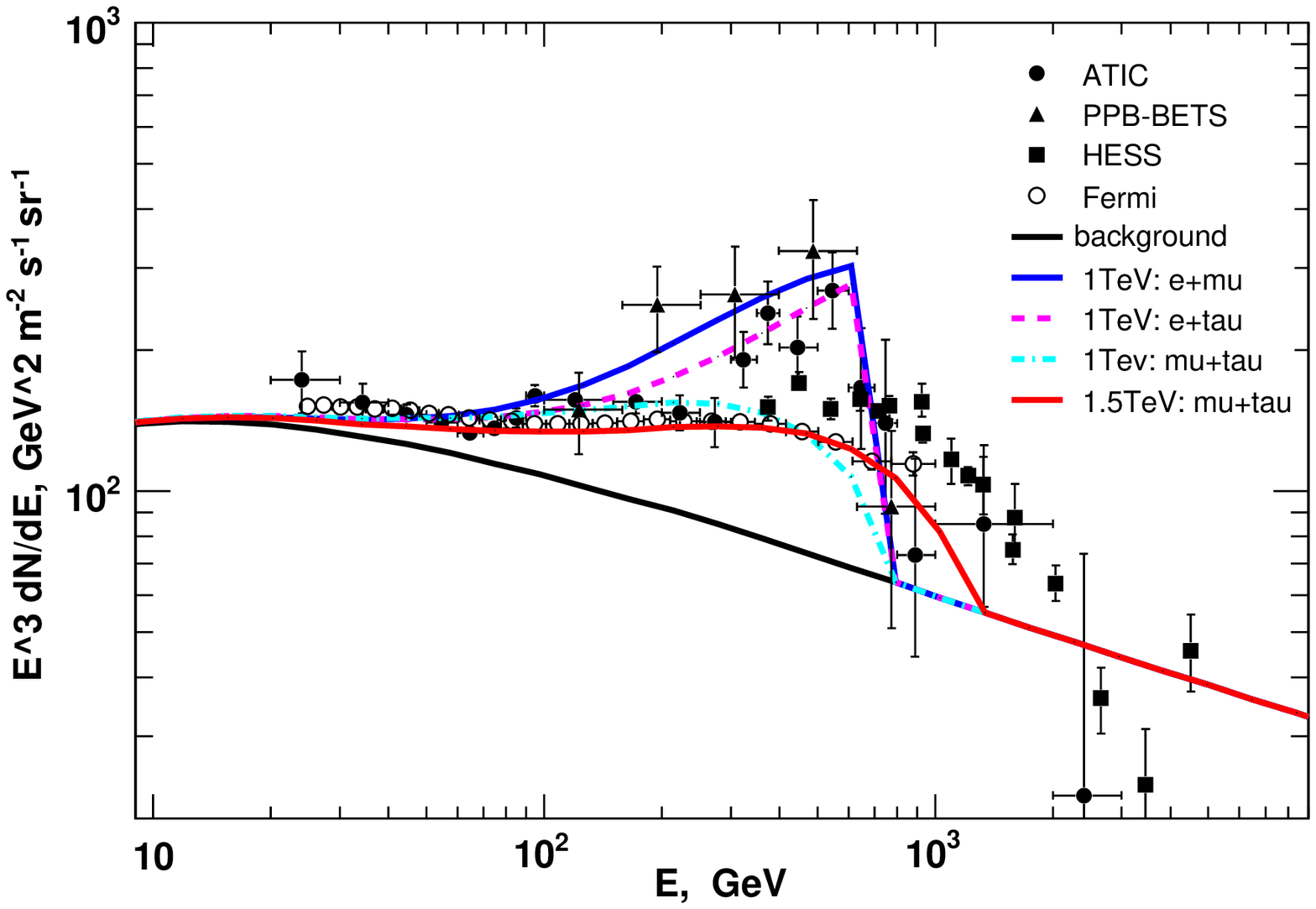}
\caption{{\it Left:} positron fraction $e^+/(e^++e^-)$ predicted
in the $U(1)_{L_i-L_j}$ model compared with the observational data
from PAMELA \cite{Adriani:2008zr} and HEAT
\cite{Barwick:1997ig,Coutu:2001jy}. {\it Right:} the total
electron spectrum of the model, compared with observations of ATIC
\cite{Chang:2008zz}, PPB-BETS \cite{Torii:2008xu}, Fermi
\cite{fermi} and H.E.S.S. \cite{HESS:2008aa}.}\label{elec}
\end{figure}

In Fig. \ref{elec} we show the model predictions on the
$e^+/(e^++e^-)$ fraction and $e^++e^-$ fluxes together with the
observational data. The background is also calculated using
GALPROP package \cite{galprop} with the diffusion + convection
model parameters developed in Ref. \cite{Yin:2008bs}. Following
Ref. \cite{Zhang:2008tb}, to fit ATIC and PAMELA data we adopt the
DM mass $\sim $ 1 TeV and Merritt density profile. In addition for
the $U(1)_{L_\mu - L_\tau}$ (model c) we also plot the spectrum
with DM mass $1.5$ TeV to fit the Fermi result. A boost factor
$\sim 1200$ (we have included the branching ratio into neutrinos
which do not produce electron/positron excess)\footnote{Note that
the boost factor in Ref. \cite{Zhang:2008tb} is defined as
${\langle\sigma v\rangle}/{3\times 10^{26}{\rm\ cm}^3 {\rm \
s}^{-1}}$, instead of the one $\langle\sigma
v\rangle/\langle\sigma v\rangle_{x=20}$ as shown in Fig.
\ref{boost}.}, or equivalently $\langle\sigma v\rangle\approx
3.6\times 10^{-23}$ cm$^{3}$ s$^{-1}$, is found to give good
description to the data \cite{Zhang:2008tb} (The cross section for
$1.5$TeV DM to fit Fermi requires the a slightly larger value of
$\sim 5.4\times 10^{-23}$ cm$^{3}$s$^{-1}$). Considering the
errorbars of the ATIC data, we find that $\langle\sigma v\rangle=
2.7-4.5\times 10^{-23}$ cm$^{3}$ s$^{-1}$ (corresponding to a
boost factor $900-1500$) can be consistent with the observations.

For our model scenarios, in the cases a) and b) one of the pairs
is directly annihilated into electron/positron pair and another
pair with secondary electron/positron pair. As can be seen from
Fig. \ref{elec}, these two cases predict a sharp falling of
electron/positron in the
spectrum at energy about 600 GeV and fit the ATIC data very well. We
also note that the case b) is slightly favored over case a), but
the difference is very small.

For the case c), the electron/positron pairs come from secondary
decay of muon pair and tauon pairs. If one normalizes the boost
factor to PAMELA data, there is no sharp falling in the
electron/positron spectrum at around 600GeV.  This model cannot
fit ATIC data well. Since this model has a softer spectrum, one
wonders whether it can fit Fermi data. It is clear from Fig.
\ref{elec} that for DM mass of $\sim 1$ TeV it can not fit the
Fermi data either. However, we see from Fig. \ref{elec} that with
a $1.5$ TeV DM mass the soft electron spectrum can fit the Fermi
data very well, and at the same time it can also account for the
PAMELA data of positron ratio.

\begin{figure}[!ht]
\includegraphics[scale=0.5]{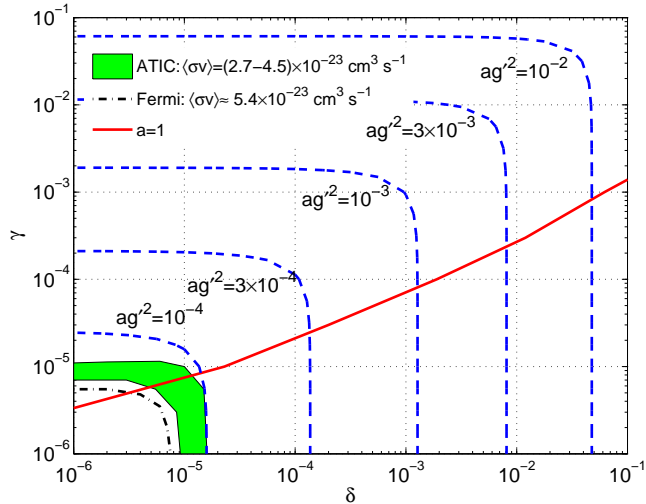}
\caption{Constraints on the model parameters from the DM relic
density and the cosmic ray data on the $\gamma-\delta$ plane.
Dashed lines show the isolines with $ag^{\prime 2}$ adjusted to
satisfy the DM relic density. Shaded region shows the allowed
parameter ranges with additional PAMELA/ATIC constraints, while
the dot-dashed line shows the constraint from the best fitting
cross section for PAMELA/Fermi data. The $a=1$ curve is given by
requiring correct relic density. }\label{para}
\end{figure}

Finally combining the constraints on the model parameters from the
relic density and PAMELA, ATIC/Fermi data we can derive the model
parameters, which are shown in Fig. \ref{para}. By properly
adjusting the $ag^{\prime 2}$, the whole parameter space of
$\gamma-\delta$ can give the right relic density. However, to give
the needed boost factor which can account for the cosmic ray
electron/positron data, the parameters are further limited in a
much narrower range, as shown by the shaded region in Fig.
\ref{para}, which gives a fit to the PAMELA/ATIC data. For the
model to fit PAMELA/Fermi data, we need a heavier DM mass and a
bit larger cross section. The constraint is shown by the
dot-dashed curve in Fig. \ref{para}. We did not take the errors of
Fermi data into account since they have very small statistic
errors. We also plot an $a=1$ curve which gives correct relic
density while setting $a=1$. The overlap between the $a=1$ curve
with the shaded region gives the correct relic density and boost
factor to account for ATIC and PAMELA data with universal gauge
couplings between dark matter and SM leptons. The allowed range of
$g'$ with $a=1$ is $6.5\times 10^{-5} \lesssim g'^2 \lesssim
9.2\times 10^{-5}$. Then we have $ 5.5\times 10^{-6} \lesssim
\gamma \lesssim 7.5\times 10^{-6}$ and $5\times 10^{-6}\lesssim
\delta \lesssim 1.2\times 10^{-5}$ for $a=1$ to give correct relic
density and boost factor. Similarly for the model to fit
PAMELA/Fermi data we have $g'^2\approx 5.8\times 10^{-5}$,
$\gamma\approx 4.8 \times 10^{-6}$ and $\delta\approx 3.5\times
10^{-6}$. In principle there should also be a narrow range satisfy
the experimental constraints, which are not shown here. For $a\neq
1$ case, one can extract $a$ and $g^{\prime 2}$ separately from
Fig. \ref{elec} by the fact $\gamma \approx 3 g^{\prime 2}/12 \pi$
in our models with a known $a g^{\prime 2}$. An interesting thing
to note is that the $Y^\prime$ charge of the dark matter does not
need to be very different from the SM leptons.

\section{Discussions and Conclusions}

In this paper we studied a class of leptophilic dark matter model
with $U(1)_{L_i - L_j}$ gauge interactions to account for the
recent cosmic ray results. We have shown that two of the anomaly
free gauged $U(1)_{L_i - L_j}$ models, the gauged $U(1)_{L_e -
L_\mu}$ and $U(1)_{L_e - L_\tau}$ models, can naturally explain
the excess in electron/positron spectrum at ATIC,  while another
model with  $U(1)_{L_\mu - L_\tau}$ gauge coupling can naturally
account for the electron spectrum at Fermi. We have seen that in
order to fit data the regions allowed for $g^{\prime 2}$ and $a$
are constrained to narrow regions. Plus the facts that the dark
matter is required to be about 1 TeV to explain the sharp falling
in electron/positron excess at ATIC or about 1.5 TeV to best fit
the Fermi and HESS data, and $Z^\prime$ mass is required to be
about two times of the dark matter mass to have a large boos
factor via the Breit-Wigner enhancement mechanism, the models
parameters are all determined in narrow regions.

Since the model parameters are restricted to narrow regions,
one has to check if the constraints on the models
are compatible with other processes. A closely related process is direct dark
matter search. The annihilation process, $\psi + \bar \psi \to e^+
+ e^-$, producing the relic dark matter density and the
electron/positron excess, is related to the direct detection
process by changing the interaction from s-channel to t-channel,
i.e., $\psi + e \to \psi + e$. One may worry if the enhanced
s-channel cross section, a large boost factor, will also lead to a large t-channel cross
section resulting in conflict with direct detection results.
This is not the case.

For $\psi + e \to \psi +e$ collision with $\psi$ mass of around 1
TeV ( or 1.5 TeV), the energy transferred to the electron is very
small, at the eV order, far below the threshold energy of present
detectors. However, if the dark matter particle $\psi$ kicks out a
tightly bound electron, the energy transferred to the electron may
be larger and detectable. But even this happens, the interaction
cross section for the t-channel is still too small because in this
case there is no Briet-Wigner enhancement, $\sigma(\psi + e \to
\psi + e) = (6 a^2 g^{\prime 2}/\pi)m^{*2}_e/m^4_{Z^\prime}$. Here
$m^*_e$ indicates a free electron or a tightly bound electron.
Using the bounds on the model parameters obtained before, we find
that the cross section is of order $(m^{*2}_e/\mbox{GeV}^2)\times
10^{-48}\mbox{cm}^2(2\mbox{TeV}/m_{Z^\prime})^2$, which is well
below the sensitivities of the present detectors.

Low energy processes such as g-2 of electron, muon or tauon,
$e-\nu$ collision data, and also high energy experiments at  LEP
have put stringent constraints on possible electron interaction
with new gauge particles. However, in contributions from
$Z^\prime$ are all proportional to $g^{\prime 2}/m^2_{Z^\prime}$
 which is of order
$10^{-11}/a$(GeV$^{-2})(2\mbox{TeV}/m_{Z^\prime})^2$ . If $a$ is
of order one there is no conflict with data.

Direct production of $Z^\prime$ will be very difficult at LHC
since at tree level the $Z^\prime$ only interacts directly with
electron, muon/tauon. ILC will not be able to produce directly
$Z^\prime$ either since its energy is below the $Z^\prime$ mass.
But $Z^\prime$ can be produced at CLIC for models a) and b) where
the $e^+e^-$ center of mass frame energy can be as high as $3$
TeV. It may be interesting to scan the machine energy around the
$Z'$ resonant region to discover it. The decay widths of $Z'$ in
models considered here are of order 10 MeV. This requires a very
careful scanning. For model c), a muon collider may provide a
chance to produce the relevant $Z^\prime$ boson.

Finally we point out that our study can be easily extended to a
larger group of dark matter models that are leptophilic. Once we
assume the Breit-Wigner enhancement takes effect to explain both
the relic density and positron excess today the model parameters
can be determined within a quite definite range, which should be
similar to the present results.

 In summary, in the present work we have constructed simple
dark matter models with gauged $U(1)_{L_i-L_j}$ interactions. We
have shown that the direct electron component in the annihilation
final states are necessary to account for the sharp falling at the
ATIC electron spectrum at $\sim 600$GeV. The $U(1)_{L_e-L_\mu}$
and $U(1)_{L_e-L_\tau}$ models with the dark matter mass around 1
TeV  are in excellent agreement with ATIC data. However, to
explain Fermi data which does not show the sharp falling indicated
by ATIC data, one should avoid having direct electron component.
This fact favors $U(1)_{L_\mu - L\tau}$ model. All these models
predict quite definite coupling constant, $Z^\prime$ mass and
width, which are consistent with all present collider and other
low energy data.

At present ATIC or Fermi data is consistent with PAMELA data
separately, but ATIC and Fermi data do not agree with each other.
Although one can find models which fit one of the two data sets,
i) PAMELA and ATIC, and ii) PAMELA and Fermi, it is obvious that
these models cannot simultaneously all be correct. To finally
decide which DM model is correct, experimental data have to give
an unique set of data which can only be achieved by future
improved experiments. Only by then one can decide the correct dark
matter model.

\acknowledgements We thank Wanlei Guo and Henry Wong for useful discussions and Juan Zhang for help plotting one figure.
This work was supported in part by the NSF of
China under grant No. 10773011, by the Chinese Academy of
Sciences under the grant No. KJCX3-SYW-N2, by NSC and NCTS.


\end{document}